\documentclass[aps,prl,amsmath,amssymb,reprint,superscriptaddress]{revtex4-2}

\usepackage{graphicx}
\usepackage{xcolor}
\usepackage[normalem]{ulem}

\begin{document}

\title{Resurgent revivals in bosonic quantum gases: a striking signature of many-body quantum interferences}
\newcommand{\LiegeUniversity}{CESAM research unit, University of Liege, 4000 Li\`ege, Belgium}
\newcommand{\Steve}{Department of Physics and Astronomy, Washington State University, Pullman, WA USA}
\newcommand{\LPTMS}{Université Paris-Saclay, CNRS, LPTMS, 91405, Orsay, France}

\author{Peter Schlagheck}
\email{Peter.Schlagheck@uliege.be}
\affiliation{\LiegeUniversity}
\author{Denis Ullmo}
\affiliation{\LPTMS}
\author{Gabriel M.~Lando}
\affiliation{\LPTMS}
\author{Steven Tomsovic}
\affiliation{\Steve}

\begin{abstract}
Matter wave revivals depend on a delicate interplay of constructive many-body quantum interferences in the developing dynamics of an ultracold bosonic system in an optical lattice.  It is shown that the interplay between weak intersite tunneling and strong onsite interactions can lead to the quantum dynamics of a density wave displaying several features not found in the mean-field limit: occupancy oscillations, resurgent revivals, and a (anti-) synchronization of revival peaks and occupancy oscillation peaks.  This implies cooperative interference effects that alternate between constructive and destructive features leading to the peak revival behaviors. These many-body quantum interference phenomena create striking features in various observables, which are accessible in experimental measurements.
\end{abstract}



\maketitle

The quantum dynamics of isolated, far-from-equilibrium, strongly-interacting many-body systems exhibit  an extensive range of phenomena intimately linked to developing quantum interferences and superposition.   They underlie entanglement generation~\cite{Islam15}, relaxation towards equilibrium~\cite{Kaufman16}, the saturation of out-of-time-orderded correlators~\cite{Rammensee18}, other post-Ehrenfest time scale phenomena~\cite{Tomsovic18}, and they play a significant role in the context of scarring phenomena related to the absence of thermalization~\cite{Bernien17, Turner18, Serbyn21}.   Early on, the groundbreaking experiment of Greiner et al.~\cite{Greiner02b} demonstrated the dispersal and subsequent revivals of matter wave fields using a Bose-Einstein condensate in an optical lattice.  This cannot be described in a mean field approximation, such as the truncated Wigner approximation (TWA)~\cite{Steel98, Sinatra02, Polkovnikov10}, due to the necessarily elaborate constructive many-body quantum interferences required to reconstruct the matter wave field.

Revivals are a well known and remarkable signature of quantum interferences in the context of ordinary Schr\"odinger dynamics~\cite{Bluhm96, Robinett04}.  Experimentally realized with electronic wave packets in Rydberg atoms, both revivals~\cite{Yeazell90} and fractional revivals~\cite{Yeazell91} have been observed.  There the evolution of minimum uncertainty wave packets initially disperse just as a consequence of Heisenberg's uncertainty principle and underlying nonlinear classical dynamics.  For a wave packet initially centered somewhere along a periodic orbit, the expected recurrences collapse at the outset, the more nonlinear the dynamics, the more rapidly they diminish. Later, for integrable dynamical systems with effectively one degree of freedom or a few degrees of freedom and rationally related time scales, the wave packet is observed to reconstruct as a result of a delicate balance of multiple constructively interfering components; note however that it is possible to observe wave packet revivals even in low dimensional chaotic systems under exceptional circumstances~\cite{Tomsovic97}.

In the context of many-body bosonic systems, the perfect analogy to a minimum uncertainty wave packet is a Glauber coherent state~\cite{Glauber63}.   Nevertheless, surprisingly few studies of coherent state revivals have been reported in the context of ultracold quantum gases, which otherwise constitutes a rather powerful platform for performing an enormous variety of quantum simulations~\cite{Bloch08}.  The notable exception is the aforementioned experiment of Greiner et al.~\cite{Greiner02b}. There the corresponding ``classical dynamics'' is not related to the condensate atoms' motional degrees of freedom, but rather to the underlying mean-field description in terms of a condensate wavefunction whose phase undergoes oscillations within each well. 

At a very elementary level their experiment can be viewed as the quantum dynamics of a collection of coherent states in independent single well Bose-Hubbard models (Kerr Hamiltonian), which implies a perfect periodic train of revivals owing to the quadratic scaling with particle number of the phase oscillation frequency (given by the condensate's chemical potential)~\cite{Greiner02b}.  However, the Bose-Hubbard model's dynamics becomes considerably richer with the introduction of tunneling between wells.  Phenomena emerge as varied as superfluid-Mott insulator phase transitions~\cite{Fisher89,Greiner02a}, Josephson oscillations and self-trapping~\cite{Smerzi97,Albiez05}, and chaotic dynamics~\cite{Kolovsky04}, to name a few of the possibilities.  It is therefore natural to investigate many-body quantum interference phenomena in bosonic matter field dynamics as a function of the competition between tunneling and interaction strengths.  In this regard, considerable effort has been directed toward quantum entanglement and thermalization studies~\cite{Lauchli08,Daley12,Alba13,Kaufman16,Pudlik13}.

Relatively speaking, far less attention has been paid to the revivals themselves; although see~\cite{Veksler15} for a semiclassical analysis in the opposite weakly interacting regime.  It turns out that even though the perfect revivals of the tunneling-free Bose-Hubbard model must degrade with increased tunneling strength, imperfect revivals nevertheless persist up to an appreciable magnitude.  Their presence evinces a significant amount of constructive many-body quantum interference over a broad dynamical range.  Under these circumstances remarkable and exotic revival behaviors emerge.  The purpose of this letter is: i) to give an account of pronounced many-body quantum interference effects in the occupancies, i.e.~simple one-body observables; ii) to investigate the coherence, the experimental measure of~\cite{Greiner02b}; iii) to illuminate a resurgent behavior of the revivals; and iv) to elucidate the working of a resonant synchronization which arises.

Consider the dynamical situation of an optical lattice with periodic boundary conditions in which alternating sites are loaded with coherent states of differing mean occupations and/or phase, i.e.~a density wave arrangement.  In the mean field description of this arrangement, there exists a ``symmetry plane'' whose dynamics can be mapped exactly onto a Bose-Hubbard dimer model~\cite{Steinhuber20}.  The dimer has also been of great interest in its own right and as a bosonic Josephson junction~\cite{Albiez05, Zibold10, Campbell20}.  It is given by 
\begin{equation}
\hat H = -J \left(\hat a_1 \hat a_2^\dagger + \hat a_2 \hat a_1^\dagger \right) +\frac{U}{2}{\Big[}\hat n_1\left(\hat n_1 -1\right) + \hat n_2\left(\hat n_2 -1\right) {\Big]}
\end{equation}
where $J$ gives the tunneling strength and $U$ the interaction strength.  It suffices to investigate the revival dynamics of a Bose-Hubbard dimer with strong repulsive interaction and a tunable hopping strength in order to illuminate the far greater range of resultant many-body quantum interference effects.  It is quite helpful to use a scaled time $\tau=Ut/(2\pi\hbar)$ as revivals occur for $\tau$ very nearly equal an integer.  The ratio of tunneling to interaction strength divided by the mean occupancy of the sites fixes the dynamics in the mean field limit.  Thus, $\gamma= 2J/(nU)$ where $n=n_1+n_2$.  Its use enables a direct comparison of systems with different $n$ as their mean field dynamics are identical.  It also serves as the natural parameter which emerges from semiclassical and quantum perturbation theories as a function of $J$.

\begin{figure}[t]
\begin{center}
\includegraphics[width=\columnwidth]{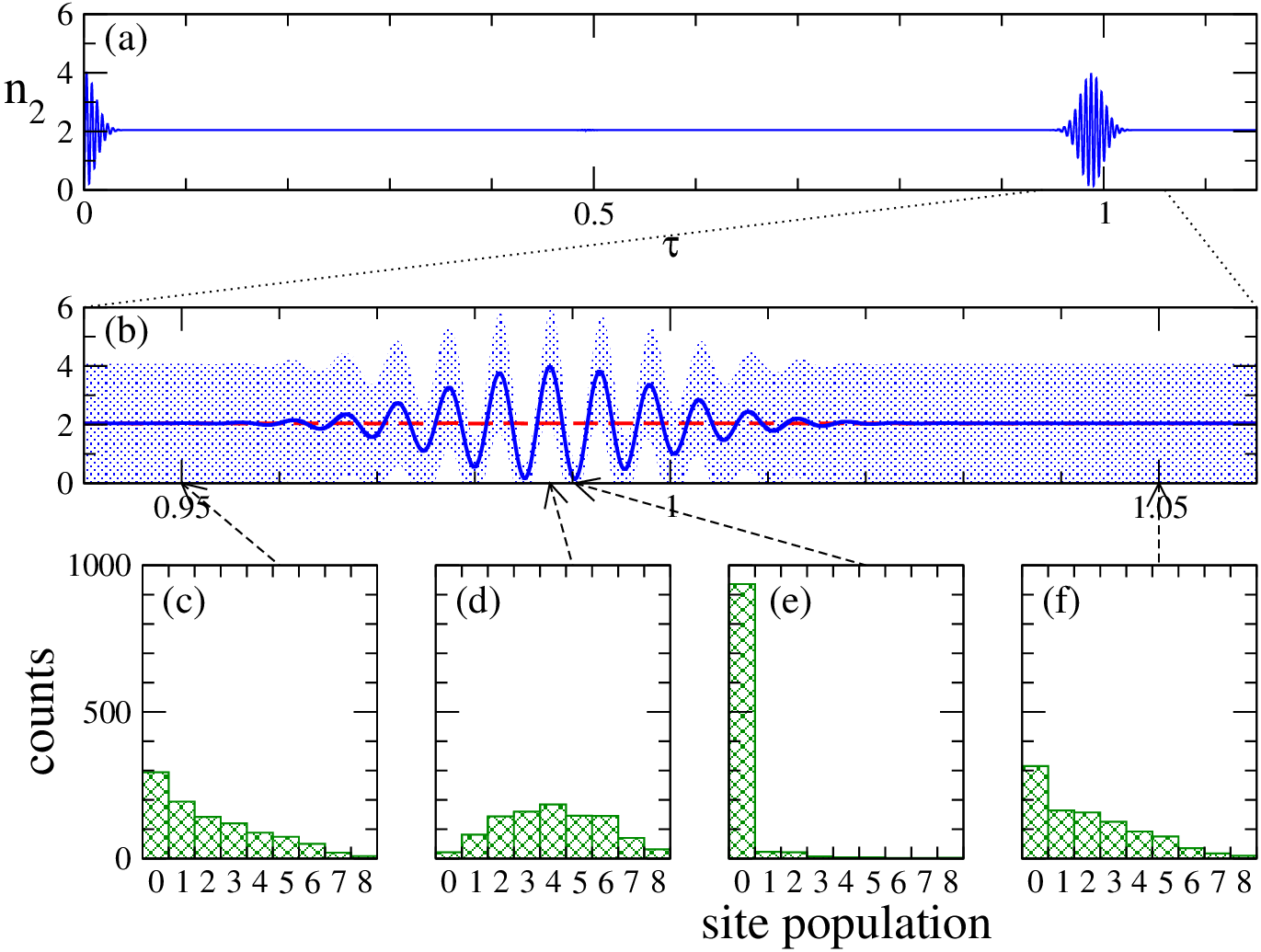}
\end{center}
\caption{(Color online) (a) Time evolution of $n_2 (\tau)$ (blue solid line), calculated for $\gamma = 0.14$ and $n = 200$. Initial population oscillations between the two sites rapidly decay, but return for $\tau \approx 1$, near the revival. (b) (zoom) the TWA (flat red dashed line) cannot reproduce this oscillatory feature.  Panels (c)-(f) display outcomes of numerically simulated occupancy measurements for 1000 repetitions. After the initial transient and well away from $\tau \approx 1$ (c),(f) have a thermal-like density.  Clear deviations are found at a local maximum (d) and minimum (e) of $n_2(\tau)$. These detection statistics are reflected in the occupancy standard deviations, blue shaded zone of (b) [$n_2 - \Delta n_2,n_2 + \Delta n_2$].  
\label{fig:pops}}
\end{figure}

Noting $|z_1,z_2\rangle$ the coherent state such that $\hat a_i|z_1,z_2\rangle = z_i$, and thus $n_i = \vert z_i\vert ^2$ ($i=1,2$), the dynamics display a number of striking quantum interference effects for an initial state given by the limit of one unoccupied site, $|z,0 \rangle$, i.e.~$(n_1, n_2)=(n, 0)$.  The first observation is that the interference effects are strong enough to be clearly visible in a simple and measurable one-body observable~\cite{Kaufman16}, namely $n_2(\tau)=\langle \psi(\tau)| \hat n_2 |\psi(\tau) \rangle$; see Fig.~\ref{fig:pops}.  In the self-trapping dynamical regime, the originally empty site occupancy, $n_2(\tau)$, displays a short time transient rapid oscillation and then settles down to its long time average.  This behavior is captured properly by a TWA calculation as expected.  However in the neighborhood of the revival time, but slightly earlier, a rapid oscillation reappears that is entirely missed by the TWA.   In addition, for most times, where there are no oscillations in $n_2(\tau)$, there is a stable probability density (relative frequency) of finding a particular integer value when measuring $\hat n_2$, the most probable value being zero; the density appears roughly exponential.  Inside the time envelope of the reviving rapid oscillations, near a minimum the measurement probability is more or less a Kronecker $\delta_{{n_2}0}$.  A very short time earlier though (a fraction of $\tau/n$), the measurement probability is peaked near $4$ and has significant probability up to about $8$, thus displaying wild swings in the occupancy probabilities.  As $\gamma$ increases, the magnitude of the oscillations increase, i.e.~increasing tunneling leads to greater constructive many-body quantum interference in $n_2(\tau)$, perhaps up to $\gamma \sim 0.2$.  A good measure to illustrate the stability of the probability density is the standard deviation, $\pm \Delta n_2$.  Its range is indicated by the blue shaded region of panel (b).  Away from the reviving $n_2(\tau)$ oscillations, the probabilities are stable and little is changing, but in the oscillating region $\Delta n_2$ is rapidly varying between nearly vanishing and the long time average of $n_2(\tau)$.

Various quite different theoretical approaches can be used to analyze successfully these behaviors and those displayed in the following figures as well.  A modified quantum perturbation theory that goes up to second order in the eigenvalues, but to higher orders in the eigenfunctions, captures the oscillations and, for example, predicts that the peak of the oscillation envelope of $n_2(\tau)$ shifts forward in time by $\gamma^2/2$ from $\tau=1$, consistent with Fig.~\ref{fig:pops}.  Another approach bootstraps EBK quantization information into the TWA to create a hybrid theory, which also predicts the oscillations and $\gamma^2/2$ shift, and gets increasingly precise as the total number of particles increases.  The details of both approaches are left to a complete accounting elsewhere~\cite{Schlagheck22}.  

A second important observable, measured by Greiner et al.~\cite{Greiner02b}, is the coherence given by $a^2(\tau)=|\langle \psi(\tau) | \hat a | \psi(\tau) \rangle|^2$.  It is much more closely related to a revival by virtue of the fact that coherent states are eigenstates of $\hat a$, i.e.~$\hat a | z \rangle = z | z \rangle$.  A desirable feature is that $a^2(\tau)$ only approaches the value of $n=|z|^2$ in the case that $|\psi(\tau)\rangle$ approaches $|z\rangle$ to within some global phase.  Excluding the special case of $z=0$, renormalizing the coherence by $n$ gives a measure bounded from above by unity that only attains unity in the case of $|\psi(\tau)\rangle = |z\rangle$.  For the density wave arrangement, it turns out that the information contained in the coherence of the initially empty site behaves quite differently than that of the initially occupied site. 

\begin{figure}[t]
\begin{center}
\includegraphics[width=\columnwidth]{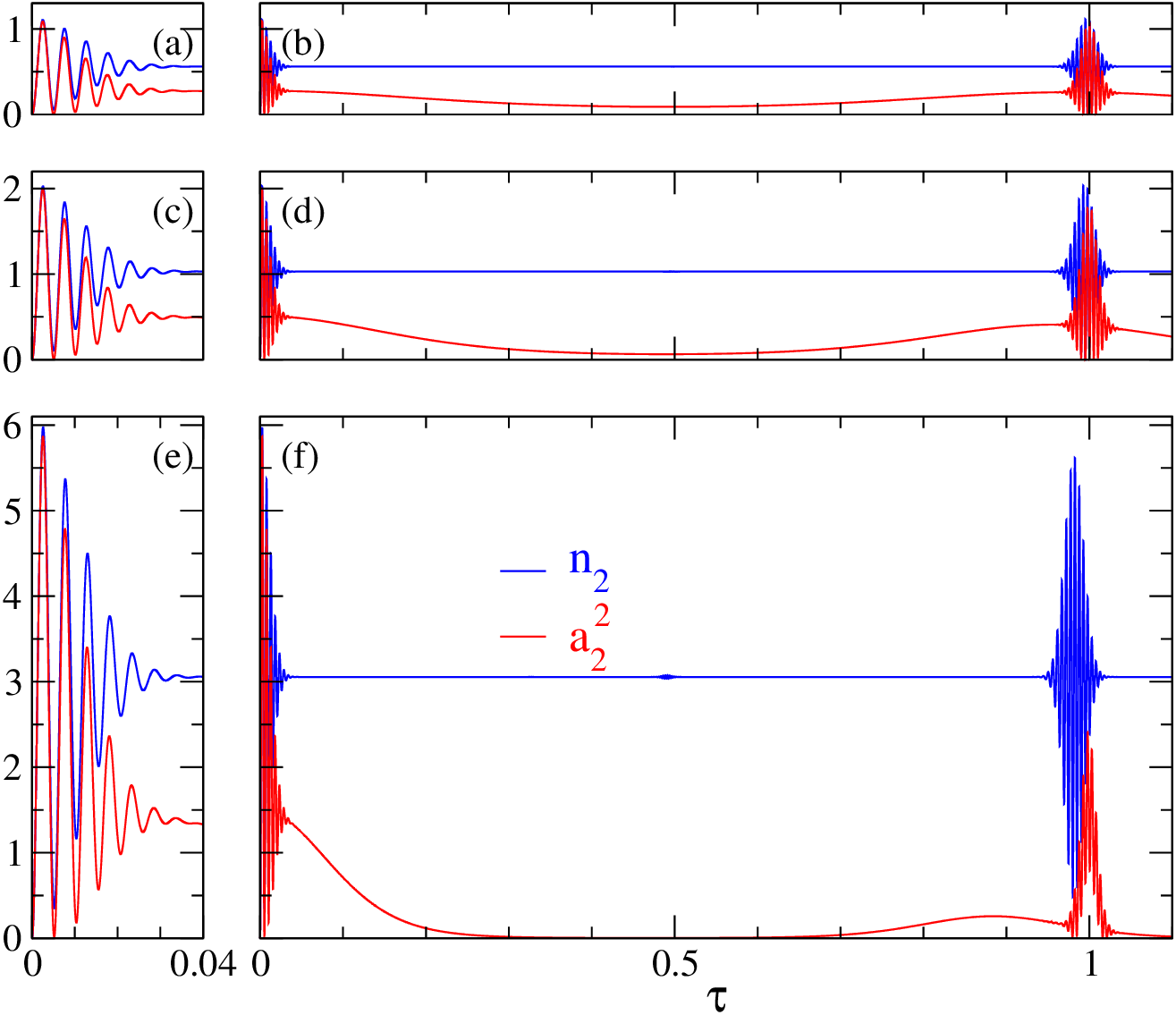}
\end{center}
\caption{(Color online) Time dependent mean occupancy $n_2(\tau)$ (blue line) and coherence $a^2_2(\tau)$ (red line) for $N = 200$. For panels (a),(b) $\gamma = 0.074$; (c),(d) $\gamma = 0.098$; and (e),(f) $\gamma=0.17$. As shown in the panel zooms (a),(c),(e), $n_2(\tau)$ and $a^2_2(\tau)$ initially display synchronous oscillations that rapidly decay.  Whereas the occupancy saturates at a steady value, the coherence vanishes at some point.  Later, it revives near $\tau \simeq 1$ (see Fig.~\ref{fig:sync} for a zoom on the revival feature).  
\label{fig:coh}}
\end{figure}

Beginning with the unoccupied site, for the sake of better illustration both $a_2^2(\tau)$ and $n_2(\tau)$ are plotted together in Fig.~\ref{fig:coh} versus $\tau$ [without renormalizing by $n_2(\tau)$] for a range of $\gamma$ values.  In the initial decay of oscillations, it is seen that the two measures begin approximately equal indicating that the state remains close to a coherent state unentangled with the other occupied site.  After a couple oscillations or so, $a_2^2(\tau)$ decreases more and more relative to $n_2(\tau)$ and after the oscillations die down it slowly vanishes.  This decay rate increases with increasing $\gamma$.  In the region of the revived oscillations, it returns close to the value of $n_2(\tau)$ again for an oscillation or so.  However, once $\gamma$ is sufficiently large that the long time average of  $n_2(\tau)$ exceeds unity, $a_2^2(\tau)$ stops recovering to a nearly perfect coherent state, and only partially succeeds in doing so.   Similarly, the revived $n_2(\tau)$ oscillations cease reaching all the way down to zero beyond the same value of $\gamma$.  Also, as $\gamma$ increases there is an increasing shift between the two envelopes of the $a_2^2(\tau)$ and $n_2(\tau)$ oscillations.  The envelope of $a_2^2(\tau)$ shifts almost imperceptively later by $3\gamma^2/(2n)$ whereas the $n_2(\tau)$ envelope shifts earlier by approximately $\gamma^2/2$, as previously mentioned.

\begin{figure}[t]
\begin{center}
\includegraphics[width=8.5 cm]{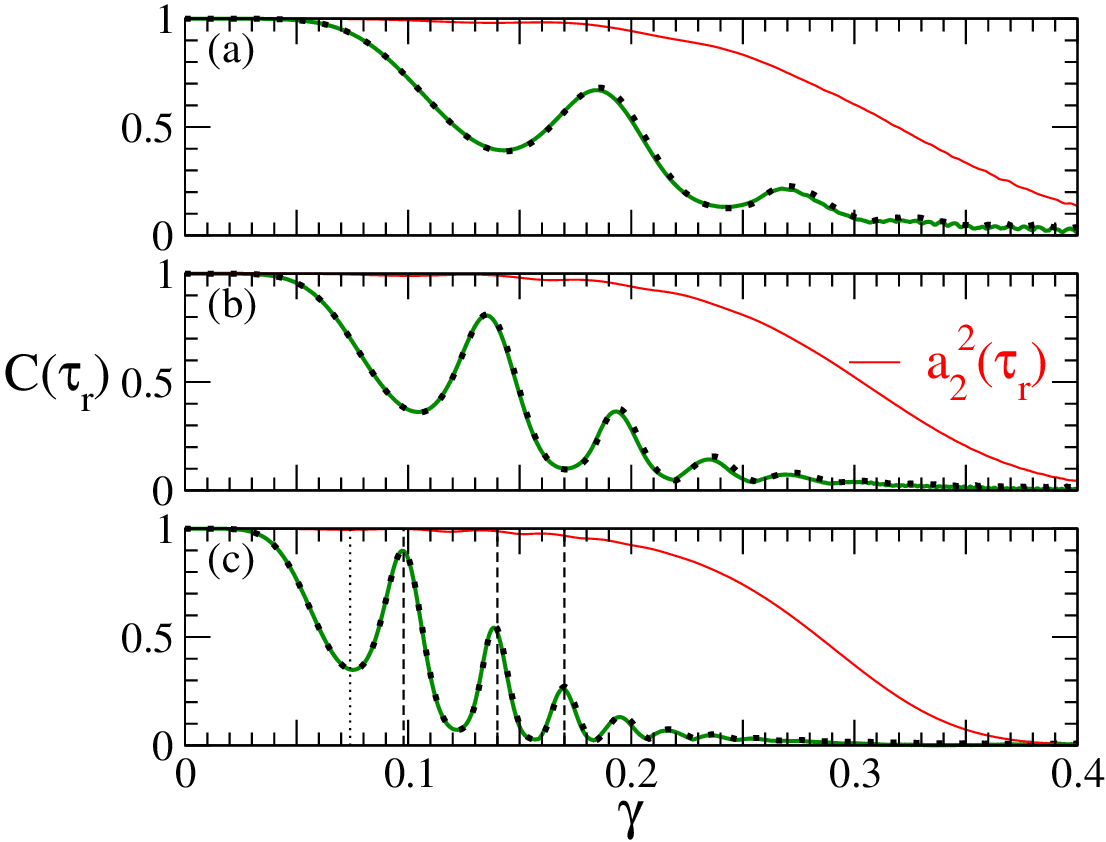}
\end{center}
\caption{(Color online) Maximal value of the revival peak for the renormalized occupied-site coherence (red line) and the return probability to the initial state (green line) as a function of $\gamma$, for (a) $n = 50$, (b) $n = 100$, (c) $n = 200$. Note the pronounced variations in the return probability peak heights. On panel (c), the  dashed lines at $\gamma = 0.098, 0.14$, and   $0.17$ indicate local maxima  of the curve associated with the return probability.  As seen on Figs.~\ref{fig:coh} and \ref{fig:sync}, these local maxima correspond to values of $\gamma$ for which the long time average of $n_2(\tau)$ takes on an integer value; see Fig.~\ref{fig:coh}.  The local minima (dotted line at $\gamma = 0.074$)
are found at $\gamma$ values for which the long time average of $n_2(\tau)$ is half-integer; see again Figs.~\ref{fig:coh} and \ref{fig:sync}.  The revival peak heights are essentially perfectly reproduced by a high order quantum perturbative calculation indicated by the black dots. Similar, but far smaller variations of the renormalized occupied-site coherence exist, but are effectively imperceptible in this figure. 
\label{fig:revival}}
\end{figure}

In the context of single particle physics, such as pump-probe experiments in Rydberg atom electronic wave packets~\cite{Yeazell90, Yeazell91}, the return probability (autocorrelation function), ${\cal C}(\tau) = \left|\langle \Psi | \Psi(\tau) \rangle\right|^2$, is the measure most often considered experimentally and theoretically~\cite{Robinett04}.  A good measure for the quality of a revival is given by its peak value.  Unity indicates a perfect revival.  This is similarly true for the renormalized coherence for the occupied site, $a_1^2(\tau)/n$.  In Fig.~\ref{fig:revival}, the peak values of both measures are plotted as a function of $\gamma$ for various total particle numbers; from top to bottom, the mean particle numbers $n=50$, $100$, and $200$, respectively.  The occupied site coherence shows that at its peak, the occupied site coherent state is nearly perfectly reconstructed through many-body quantum interferences in its time evolution for $\gamma$ values up to approximately $\gamma =0.25$.  Beyond this it decays more and more quickly with increasing particle number.  The robustness of the coherence is somewhat surprising as this dynamics, $\gamma =0.25$,  is halfway towards the self-trapping transition at $\gamma=0.5$ for the initial state.  It therefore represents a rather strong perturbation to the dynamics, and revivals, which depend very sensitively on delicate constructive interferences, are typically easily destroyed. 

\begin{figure}[h]
\begin{center}
\includegraphics[width=\columnwidth]{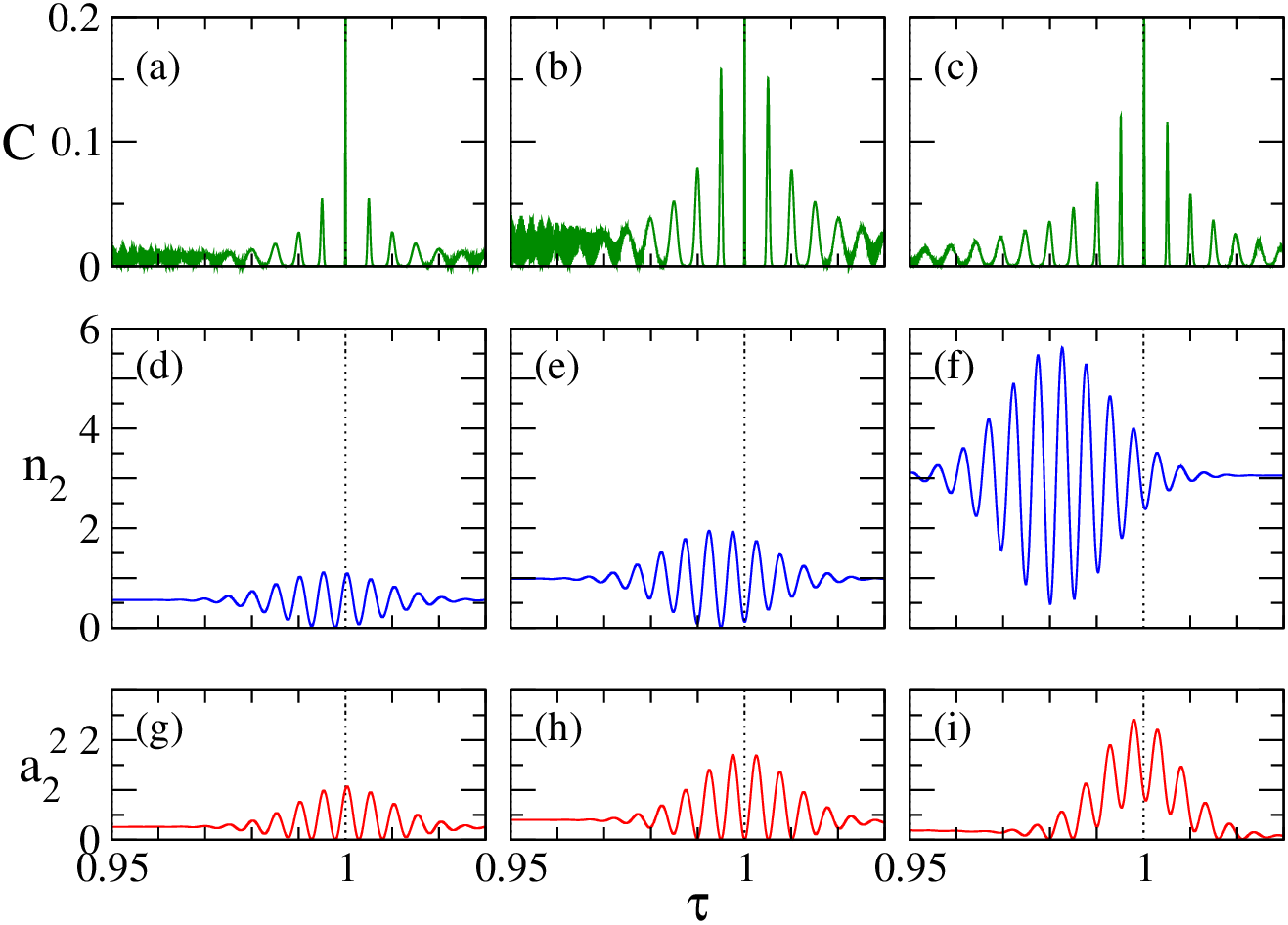}
\end{center}
\caption{(Color online)
  Zoom on the revival feature (a-c) in the return probability to the initial state, (d-f) in the occupancy on site 2, and (g-i) in the coherence on site 2, for the parameters $n = 200$ and (a,d,g) $\gamma = 0.074$, (b,e,h) $\gamma = 0.098$, and (c,f,i) $\gamma=0.17$. As seen in Fig.~\ref{fig:revival}, the case $\gamma = 0.074$ (a,d,g) corresponds to a local minimum in the peak return probability as a function of $\gamma$, whereas the other cases shown here correspond to local maxima in the return probability.  In the case of such a local maximum, the occupancy on site 2 becomes minimal at the instance where the return probability revival takes place (vertical dotted lines), thereby maximizing the population on site 1, whereas a maximal population on site 2 is encountered at the revival time in the case of a local minimum in the peak return probability.  
\label{fig:sync}}
\end{figure}

In a little bit of a contrast, the return probability decays more quickly since it contains information about both the occupied and unoccupied sites.  Figures \ref{fig:pops} and \ref{fig:coh} illustrate that the coherent state  character of the wavefunction on the  initially unoccupied site  degrades more rapidly with increasing $\gamma$, and therefore so must the return probability.  
Nevertheless, it shows a certain robustness against destruction with increasing $\gamma$, and it also displays large variations, being far from a monotonically decreasing function.  In a sense, the revival is exhibiting a resurgent behavior with increasing perturbation, which is, a priori, rather surprising.  Clearly, increasing the mean particle number $n$, increases the numbers of successive maxima and minima that appear in the curves.  Each local maximum represents a resurgence in the revival, whereby increasing $\gamma$ from the previous minimum has increased the revival quality and the amount of constructive interference.  Curiously, the $\gamma$ values of the resurgent revival peaks arise for integer values of the long time average of $n_2(\tau)$ and the local minima correspond to the half-integer values (compare Fig.~\ref{fig:revival}(c) with Figs.~\ref{fig:coh} and \ref{fig:sync}).  

The near perfect agreement of the quantum perturbation theory with the quantum calculations and its necessary reliance on high order mixing in the eigenstates hints that as they acquire increasing components of greater site $2$ occupancies, the resurgence in the revivals is due to some kind of a cooperative interference effect.  The unperturbed dimer starts with two time scales, the revival time, $\tau_r$, and phase oscillation period $\tau_p=\tau_r/n$.  In the canonical approach to understanding revivals~\cite{Bluhm96}, the time scale $\tau_r$ is related to the discrete second difference of eigenvalues whereas $\tau_p$ is related to the discrete first difference.  From either the quantum perturbative or the EBK-bootstrapped-TWA perspectives~\cite{Schlagheck22}, increasing $\gamma$ shifts these frequencies weakly.  For example, the revival time at first shifts roughly $\tau_r(\gamma)\approx \tau_r(0) [1+3\gamma^2/(2n)]$as $\gamma$ increases from zero.  This is the time at which the two frequencies remain as close to an integer relationship as possible.

However, there is a dependence on the $n_2$-site occupation probability distributions, and as higher occupancies are populated with increasing $\gamma$, there is a general decay in the quality of the revival and magnitude of the $n_2(\tau)$ oscillations with both completely disappearing well before the self-trapping transition is reached.  A clue to the additional mechanism responsible for the resurgence in the revivals on top of this simple picture is illustrated in Fig.~\ref{fig:sync}.  In the left most column, it is shown that the first \emph{minimum} in the peak return probability occurs for a $\gamma$ value at which the return probability peaks in time at exactly the same time as the $n_2(\tau)$ oscillation peaks.  The other columns illustrate that the resurgent revival \emph{peaks} are at $\gamma$ values for which the return probability peak is synchronized with a local minimum in the $n_2(\tau)$ oscillations.  This synchronization coincides with the long time averaged $n_2(\tau)$ taking on integer values; the minima are at half-integer values.  This (anti-) synchronization that minimizes the occupancy of site $2$ at the moment that the system is rebuilding the initial coherent state to a maximal extent in site $1$ cooperatively aids the overall constructive interference necessary for the overall revival.

In summary, the competition between intersite tunneling and intrasite particle interactions creates elaborate revival phenomena leading to occupancy oscillations, resurgent revivals, and (anti-) synchronization of occupancy and revival peaks due to cooperative many-body quantum interference effects.  All of these effects are amenable to experimental verification via an adaptation of the pioneering coherence experiment of Greiner et al.~\cite{Greiner02b}. The procedure that was carried out in the latter would have to be applied to optical superlattices~\cite{Folling07} that are effectively constituted by sequences of spatially displaced Bose-Hubbard dimers between which tunneling is suppressed. Brillouin zone mapping techniques~\cite{Folling07} can be used to displace the populations on the two sites to distinct regions in momentum space, which allows one to access the coherences associated with site 1 or 2, whereas site population statistics as shown in Fig.~\ref{fig:pops} can be obtained with quantum gas microscopes \cite{Kaufman16}. Owing to the connection between the classical dynamics of the dimer and that of the symmetry plane of an ultracold atom density wave in an optical lattice, analogs of these effects should also exist in the latter case, and it would be interesting to extend the analysis beyond the symmetry plane to understand how the other dynamical degrees of freedom impact all of these novel behaviors.

\acknowledgments

We thank Klaus Richter and Juan Diego Urbina for critical discussions. One of us (GML) gratefully acknowledges support through the Grant No. ANR-17-CE30-0024.


\end{document}